\documentclass[graybox]{svmult}

\usepackage{type1cm}        % activate if the above 3 fonts are
\usepackage{makeidx}         % allows index generation
\usepackage{graphicx}        % standard LaTeX graphics tool
\usepackage{multicol}        % used for the two-column index
\usepackage[bottom]{footmisc}% places footnotes at page bottom

\usepackage{newtxtext}       % 
\usepackage{newtxmath}       % selects Times Roman as basic font

\usepackage{caption}
\usepackage{subcaption}

\usepackage{hyperref}

\makeindex             % used for the subject index
\begin{document}

\title*{What machine learning can do for computational solid mechanics}
%\author{Name of First Author and Name of Second Author}
%e.g:
\author{Siddhant Kumar and Dennis M. Kochmann}

\authorrunning{Kumar, S., Kochmann, D. M.}

%\institute{Name of First Author \at Name, Address of Institute, \email{name@email.address}
%\and Name of Second Author \at Name, Address of Institute \email{name@email.address}}
%e.g:
\institute{
	Kumar, S.\at Department of Materials Science and Engineering, Delft University of Technology, 2628CD Delft, The Netherlands, \email{Sid.Kumar@tudelft.nl} \and Kochmann, D. M.\at Mechanics \& Materials Lab, Department of Mechanical and Process Engineering, ETH Z\"{u}rich, 8092 Z\"{u}rich, Switzerland, \email{dmk@ethz.ch}
}
%\institute{
%	Kumar, S. and Kochmann, D. M.\at Mechanics \& Materials Lab, Department of Mechanical and Process Engineering, ETH Z\"{u}rich, 8092 Z\"{u}rich, Switzerland\\ \email{Sid.Kumar@tudelft.nl, dmk@ethz.ch}
\maketitle
\begin{articlededication}
    ``Most people say that it is the intellect which makes a great scientist. They are wrong: it is character.'' (A.~Einstein) Here's to a notable scientist and great character on the occasion of his 70$^\text{th}$ birthday -- happy birthday, Peter! (D.~M.~Kochmann)
\end{articlededication}

\abstract{Machine learning has found its way into almost every area of science and engineering, and we are only at the beginning of its exploration across fields. Being a popular, versatile and powerful framework, machine learning has proven most useful where classical techniques are computationally inefficient, which applies particularly to computational solid mechanics. Here, we dare to give a non-exhaustive overview of potential avenues for machine learning in the numerical modeling of solids and structures and offer our (subjective) perspective on what is yet to come.}

\section{Introduction}
\label{sec:intro}
We are at the beginning of an exciting era in which the numerical solution of complex problems as well as the design and discovery of (meta-)materials and the solution of inverse problems are accelerated by big data and advances in machine learning (ML) strategies. Contrary to traditional approaches based on intricate theoretical insight, exhaustive simulations and/or experiments, data-driven learning methods exploit large data sets to identify otherwise unknown relations, thus creating efficient and invertible maps between input and output parameters -- such as simulation parameters and resulting outcomes, or (meta-)material structure and resulting properties. The latter particularly offers a quick identification and reduction of the design space, eventually leading to a faster turnaround of material design by demand.

\section{Material modeling}

\begin{figure}
\centering
\begin{subfigure}[b]{0.48\textwidth}
\includegraphics[width=\textwidth]{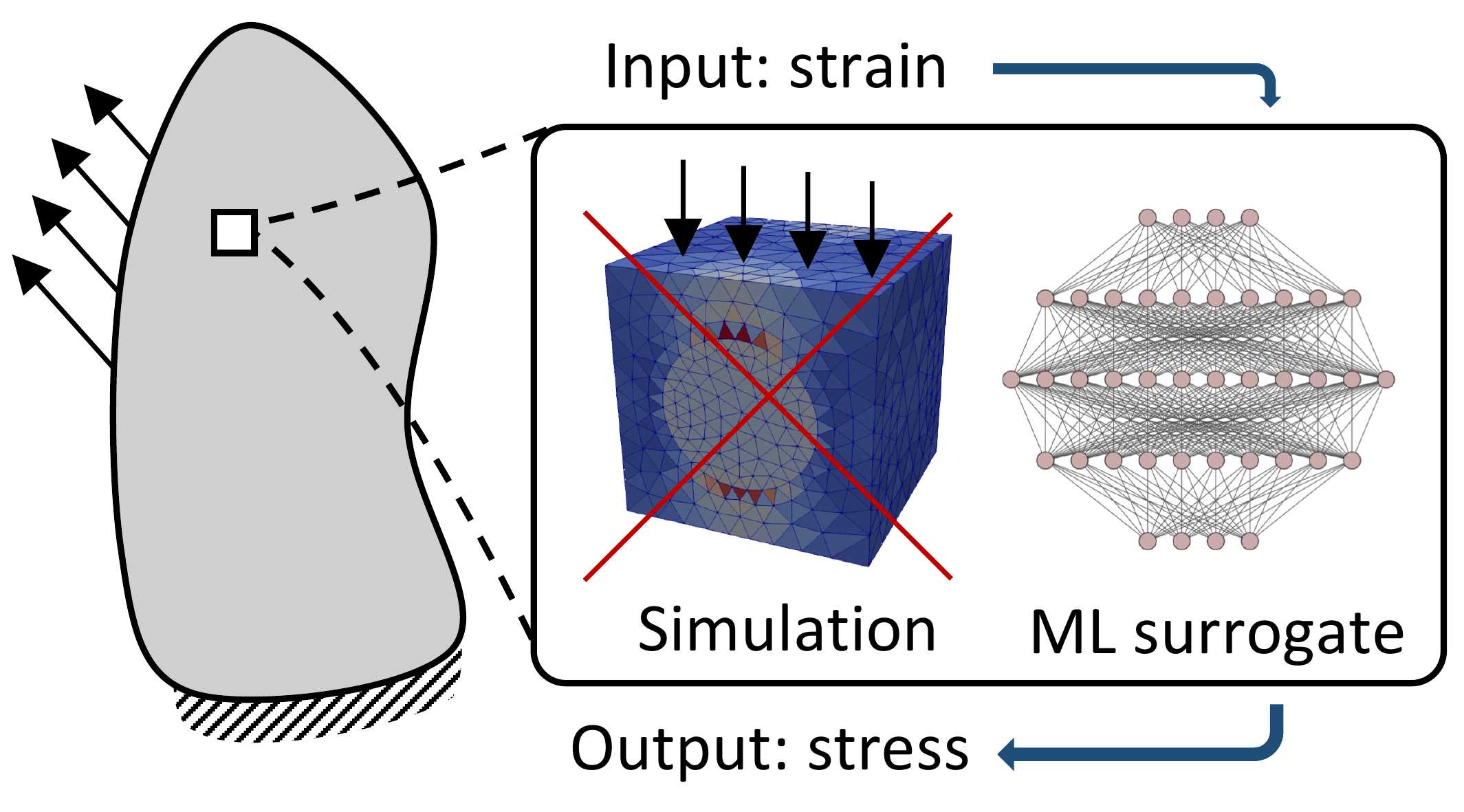}
\caption{ML-based surrogate models for bypassing microscale simulations in multiscale systems.}
\label{fig:multiscale}
\end{subfigure}
\hfill
\begin{subfigure}[b]{0.48\textwidth}
\includegraphics[width=\textwidth]{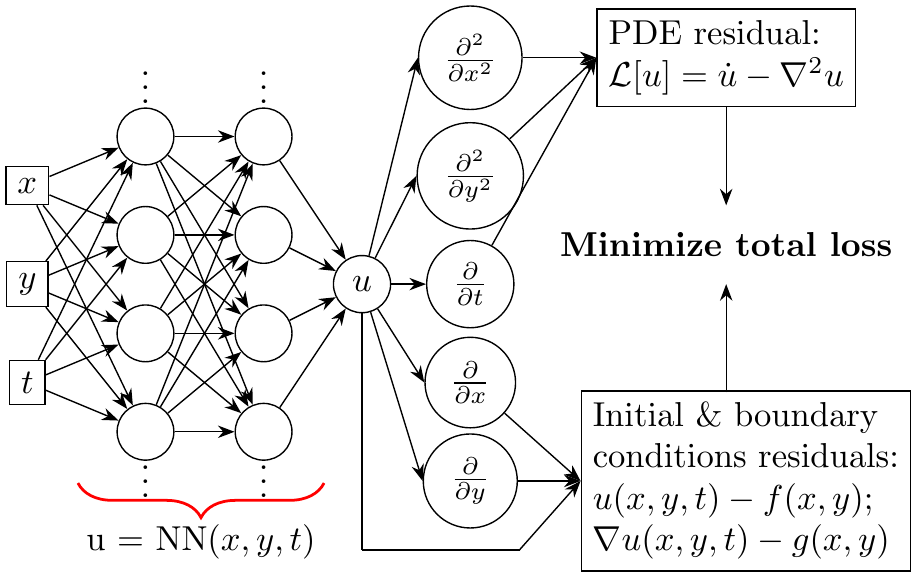}
\caption{PINNs \cite{RaissiEtAl2017_1,RaissiEtAl2019} for solving PDEs (e.g., two-dimensional heat equation shown here).}
\label{fig:pde}
\end{subfigure}
\begin{subfigure}[b]{\textwidth}
\includegraphics[width=\textwidth]{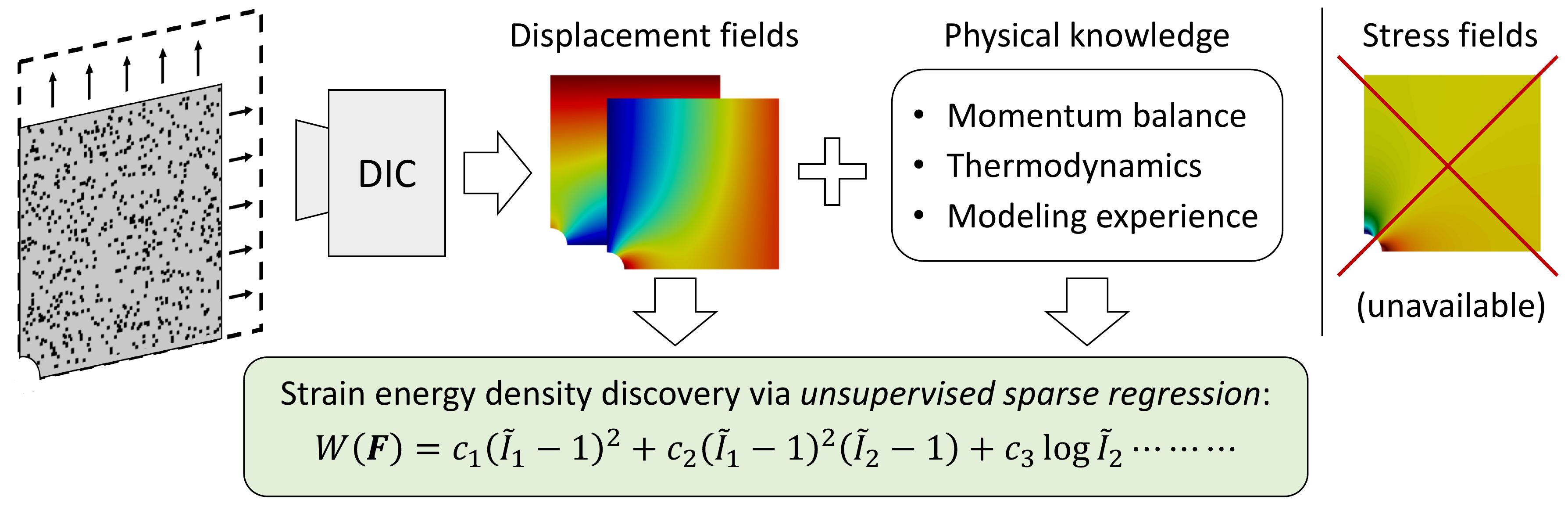}
\caption{Unsupervised discovery of interpretable and parsimonious constitutive laws using only displacement data and physical knowledge \cite{FlaschelEtAl2020}.}
\label{fig:constitutive}
\end{subfigure}
\caption{Representative examples of ML applications in material modeling.}
\label{fig:Fig1}
\end{figure}

The computational modeling of materials and structures has relied on solving century-old partial differential equations (PDEs) with ever improved numerical techniques. Especially when it comes to multiscale problems, the computational expense of simulations -- involving coupled problems on various length and/or time scales -- has remained a prohibitive challenge. ML strategies are promising for significantly accelerating not only multiscale simulations but also for identifying new constitutive models as well as for solving systems of PDEs (see Figure~\ref{fig:Fig1}).

\subsection{Accelerating multiscale simulations} \label{sec:multiscale}

Data-driven and ML-based methods have been most effective in overcoming the computational limitations of multiscale simulations by bypassing expensive lower-scale calculations and thereby accelerating macroscale simulations (Figure \ref{fig:multiscale}). Assuming a separation of scales between the microstructure and the macroscale boundary value problem (BVP), the effective constitutive response from the microscale is approximated by a pre-trained ML model to replace costly representative volume element (RVE) simulations. In computational mechanics, the focus has mostly been on \textit{surrogate models} for the homogenized microscale stress-strain  response in FE$^2$-type settings. Commonly, this has relied on supervised ML models which approximate the constitutive manifold based on training examples of stress-strain pairs. Recent work \cite{HuangEtAl2020,MozaffarEtAl2019, ReimannEtAl2019} has developed surrogate constitutive models that capture strong nonlinearity as well as path dependence (e.g., learning plastic yield surfaces for complex composites and polycrystalline metals). They were also shown to generalize well to arbitrary strain paths and are easily integrated into finite-element (FE) frameworks. Towards accelerated macroscale simulations, Capuano and Rimoli \cite{CapuanoEtAl2019} developed a general ML-driven \textit{smart finite element} method, which bypasses the need for explicitly solving internal displacement fields within an element while enforcing physical constraints. All such approaches rely on learning a model a-priori (offline) from homogenized RVE data and, consequently, require a large number of computationally expensive RVE-level simulations to obtain sufficient data for training the model. Alternatively, surrogate modeling methods can also leverage the spatial structure of the RVEs to reduce the training data requirements. Recent techniques such as graph neural networks and nonlocal pooling may take advantage of the spatial proximity of grains in polycrystalline systems to learn their constitutive response \cite{VlassisEtAl2020} and texture evolution \cite{PandeyEtAl2021} in a data-efficient manner.

\subsection{Data-driven constitutive models: beyond simulation-based training}\label{sec:constitutive}

While the above approaches rely heavily on RVE simulation data to learn the constitutive response, recent work on \textit{model-free data-driven approaches} \cite{IbanezEtAl2018,KirchdoerferEtAl2016} take a different path by using experimental data directly instead of building surrogate models. Aiming at avoiding physical-modeling biases, this results in a purely data-driven framework that searches for the closest stress-strain pair (within the dataset) which satisfies the conservation laws and compatibility conditions. These approaches have even been extended to the modeling of inelasticity \cite{EggersmannEtAl2019} and fracture \cite{CarraraEtAl2020}. Since interpolation-based modeling is generally avoided, current challenges include generalization and extrapolation capabilities to data not available in the (training) dataset, especially when dealing with high-dimensional and noisy data.
In contrast to the purely model-free philosophy is the approach of physics-informed discovery of \textit{interpretable constitutive models}. The strategy is to use experience in physical modeling to reduce the dependence on data and improve generalizability. The most popular approach has relied on sparse regression to discover a mathematically and physically interpretable form of the underlying constitutive equation \cite{BruntonEtAl2016,FlaschelEtAl2020} (in contrast to, e.g., black-box neural networks). The method involves creating an exhaustive library of mathematical functions and then sparsely selecting the combination of those which best explains the data. The motivation behind sparsity originates from the principle of Occam's razor: most physical models in nature are parsimonious in description. One of the biggest challenges in realizing data-driven constitutive models for practical application lies in material characterization; e.g., using digital image correlation (DIC) full-field strain maps become available but stress data is hard to obtain. To this end, recent work \cite{FlaschelEtAl2020} (Figure \ref{fig:constitutive}) discovered interpretable constitutive models in an unsupervised manner -- i.e., without stress data -- which presents a promising departure from curve-fitting and supervised learning based on stress data.

\subsection{Learning to solve PDEs} \label{sec:pde}

ML has also been successfully deployed to bypass the expensive numerical solution of highly nonlinear and high-dimensional BVPs. The goal is to achieve real-time solutions for a given family of BVPs with applications ranging from surgical robots to time-critical control systems. Initial attempts in this direction relied on supervised training based on thousands of BVP solutions with different boundary conditions and/or material inhomogeneity. Those ML approaches treat the inputs (boundary conditions and material distribution) and outputs (solution fields) as images and reduce the problem to image-to-image regression, using, e.g., convolutional neural networks \cite{ZhuEtAl2018}. However, such methods are data-intensive and showed limited generalization to unseen data. Additionally, the learnt model only applies to the mesh resolution of the training data.
To address those limitations,  a new class of ML methods, \textit{physics-informed neural networks} (PINNs) \cite{RaissiEtAl2017_1,RaissiEtAl2019} (Figure \ref{fig:pde}), has been proposed to solved BVPs in an unsupervised manner, i.e., using data without labels. Analogous to, e.g., the conventional Rayleigh-Ritz method, a neural network as a function of time and position is chosen as the ansatz for the solution field. Its trainable parameters are tuned to minimize the loss in the strong or weak form of the PDE and satisfaction of the boundary conditions. The major advantage of this approach is that it is unsupervised, i.e., it does not require any training dataset generated by expensive simulations. Instead, physical constrains (via the PDE-based loss) are exploited to replace the training data. Leveraging the differentiability of neural networks, this approach can also been used to calibrate parametric models of PDEs based on the observed solutions \cite{RaissiEtAl2017_2}. 
Though PINNs are successful in learning to solve BVPs, they suffer from mesh dependence of the discretization used to evaluate the PDE loss during training \cite{BhattacharyaEtAl2020}. To this end, new methods \cite{LiEtAl2020,BhattacharyaEtAl2020} are being developed for operator learning, which aims to learn the mapping in the functional space and avoid mesh dependence. In computational solid mechanics, currently both PINNs and operator learning methods have been applied only to simple PDEs in regular domains but are promising for future application to complex material systems such as those with sharp discontinuities and localization, multiscale metallic systems, or loss of uniqueness in the solutions (e.g., see \cite{KumarEtAl2020_cms,KumarEtAl2020_nme,TutcuogluEtAl2020}).

\section{Design of (meta-)materials}

Tailoring a material's microstructure to achieve targeted properties is a major challenge in natural material systems due to physical limitations. Mechanical metamaterials \cite{KochmannEtAl2019} based on random or periodic truss, plate or shell architectures overcome those limitations by letting an engineered microscale produce novel and exciting effective macroscale material properties. The natural challenge that follows is the inverse design: how can one efficiently reverse-engineer a metamaterial's design to achieve a certain properties? Traditionally, material design has heavily relied on trial and error and intuition, which is expensive and explores only a small subset of the design space. This calls for new ML-assisted computational design approaches.

\subsection{Accelerating topology optimization} \label{sec:top-opt}

\begin{figure}[t]
\sidecaption[t]
\includegraphics[width=7.5cm]{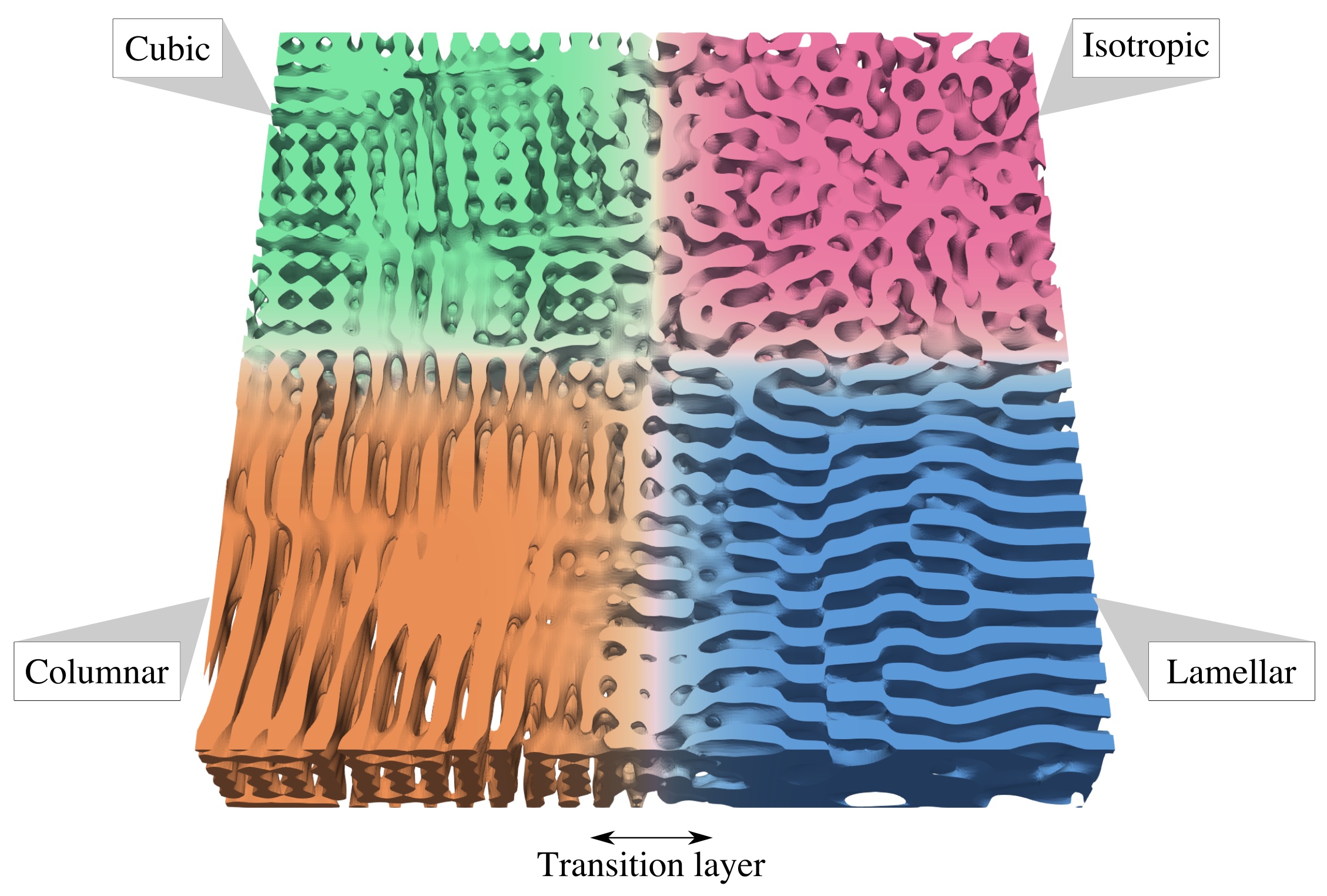}
\caption{Spindoid metamaterials, based on an approximation of structures observed during anisotropic spinodal decomposition, offer seamlessly tunable elastic anisotropy (including, e.g., cubic-, columnar-, isotropic-, and lamellar-type topologies) and functional grading \cite{KumarEtAl2020_npj}.}
\label{fig:spinodoids}
\end{figure}

\begin{figure}[t]
\sidecaption[t]
\includegraphics[width=7.5cm]{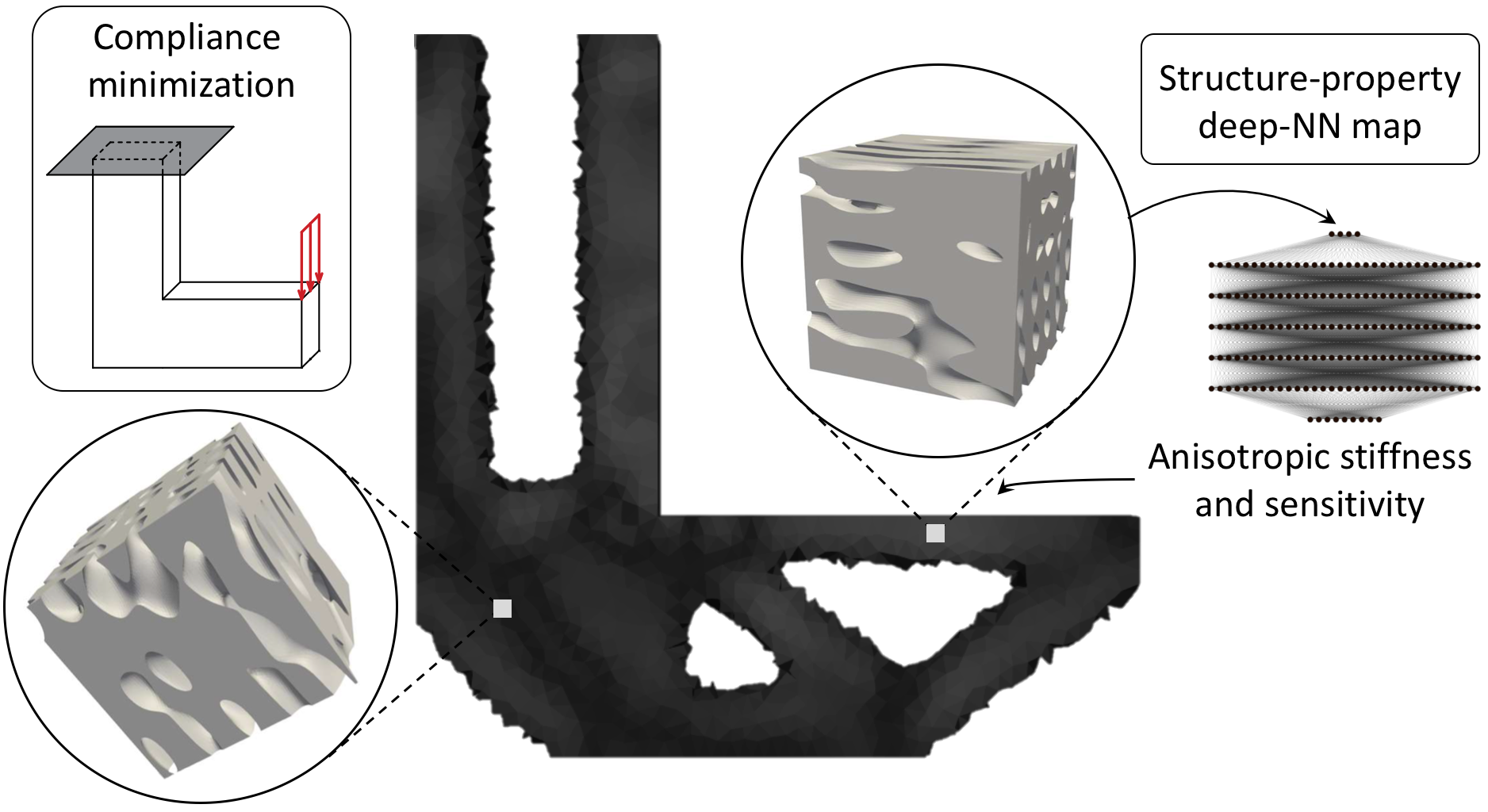}
\caption{Topology optimization of spinodoid metamaterials for minimum compliance \cite{ZhengEtAl2020_topopt}: a deep neural network provides the map from the topology to anisotropic stiffness and sensitivity.}
\label{fig:topopt}
\end{figure}

Topology optimization \cite{SigmundEtAl2013} (e.g., based on the Solid Isotropic Material with Penalization method, SIMP) has become a standard for designing structures that optimize design objectives such as minimum compliance or tailored anisotropy. This is achieved by optimizing the material distribution (solid or void) across all voxels in a discretized domain. Analogously, spatially-variant metamaterials can be designed by optimizing the design parameter(s) for each voxel. Since most metamaterials do not admit \mbox{(semi-)}analytical expressions of their homogenized responses, the latter requires expensive nested simulations based on lower-scale RVE calculations. The computational expenses are further compounded by sensitivity calculations, which rely on computing numerical derivatives (i.e., perturbing the design parameters and re-running costly RVE simulations). 
Similar to accelerating multiscale simulations, ML-based \textit{models for structure-property relations} have proven beneficial for the topology optimization of metamaterials \cite{WhiteEtAl2019,ZhangEtAl2019}. Deep neural networks are particularly advantageous in learning high-dimensional and highly nonlinear structure-property maps. They also naturally provide the exact sensitivities using automatic differentiation (which forms the backbone of neural network training by back-propagation) and avoid the computational costs and precision issues associated with numerical differentiation.  Sobolev training \cite{CzarneckiEtAl2017}, wherein sensitivity information is used for learning, can improve the accuracy, as shown, e.g., for the elastic stiffness of trusses \cite{WhiteEtAl2019}. A key open challenge arises from the topology optimization of spatially-variant structures, where prior approaches have optimized microstructural design parameters but within a limited design space since small-scale unit cells must be compatible for practical purposes to form a macroscale structure (e.g., knowing the optimal truss unit cell at each point within a macroscale body does not ensure that a smooth grating between the distinct unit cells exists). As a remedy, we recently introduced \textit{spinodoid metamaterials} \cite{KumarEtAl2020_npj} (Figure~\ref{fig:spinodoids}), which are composed of smooth bi-continuous topologies inspired by those observed in the naturally-occurring process of spinodal decomposition. Spinodoids possess a tremendous design space of anisotropic topological and mechanical properties. Their non-periodicity is a departure from periodic unit cells and avoids tessellation-related issues while admitting seamless transitions in between different anisotropic designs. This allows for combined topology optimization \cite{ZhengEtAl2020_topopt} (Figure~\ref{fig:topopt}) of local microstructural design parameters (and in turn, the local anisotropy), material distribution (solid vs.\ void), and material orientation, yielding better designs than SIMP. This is made possible by an efficient deep neural network-based surrogate model for the anisotropic design-stiffness map, which is particularly advantageous in strongly nonlinear and multiply-connected design space of spinodoids.

\subsection{Efficiently exploring design spaces} \label{sec:generative}

Exploring or searching a design space to identify optimal designs is a topical challenge both at the material level (optimizing, e.g., molecular structures for target properties) and at the structural level of metamaterials (optimizing, e.g., stiffness and strength by unit cell design). Searching a design space faces two challenges.  First, owing to its high dimensionality, the size of a design space increases exponentially with the number of design parameters, which presents computational limitations on the amount of data that can be generated for learning surrogate models of the structure-property relations. Feature selection techniques (e.g., principal component analysis and autoencoders \cite{Kramer1991}) can be used to identify the important features and reduce the dimensionality. Alternatively, the regions of the design space relevant to the design objective can be adaptively identified using, e.g., Bayesian and active learning techniques \cite{BessaEtAl2019,IyerEtAl2020}. Second, most design spaces do not admit an explicit design parameterization. E.g., a composite can be represented as a binary image, whose the pixels describe the spatial arrangement of two materials. To this end, generative ML methods such as variational autoencoders (VAE) and generative adversarial networks (GANs) have been used to learn an abstract parameterization of the design space  \cite{MaoEtAl2020,WangEtAl2020}. These are unsupervised ML methods, i.e., they do not require labeled training data and usually involve jointly training a pair of neural networks against each other. E.g., an \textit{encoder} in a VAE abstracts features from the designs and maps them to a lower-dimensional latent space where similar designs are located close to each other. The features must be sufficiently informative to allow reconstructing the design from the latent space parameterization using another neural network called the \textit{decoder}. The latent space, now representing an abstract design parameterization, can be further used as input to a regressor for surrogating structure-property relations. Alternatively, genetic algorithms and gradient-based optimization methods can be used to explore the latent space and search for the optimal design. 
Generative modeling is a promising direction for designing complex material systems. However, training of such models is usually unstable and remains as an active ML research area. From the perspective of application to \mbox{(meta-)materials}, recent works (e.g.~\cite{MaoEtAl2020,WangEtAl2020}) rely on datasets with millions of samples and are limited to two-dimensional designs.  Therefore, future improvements in the scalability of these methods are needed for application to a data-efficient design of material systems with higher-dimensional and more complex topologies.

\subsection{Inverting structure-property maps} \label{sec:inverse}

The process of developing (meta-)materials with tailored properties has traditionally followed a \textit{forward design} paradigm: computationally homogenizing (or experimentally testing) a wide range of possible structures results in an effective \textit{structure-to-property map}. Capitalizing on this strategy, methods such as topology optimization and genetic algorithms are unfortunately computationally expensive and sensitive to the initial design guess. While  ML-based generative methods such as GANs and VAEs (Section \ref{sec:generative}) reduce computational costs, they still require solving an optimization problem based on the surrogate model to achieve a targeted property, and they may converge to local optima in the design space.
A beneficial alternative is the on-demand \textit{inverse design} of optimal architectures, wherein an appropriate design is identified directly which achieves a targeted effective property. Unfortunately, identifying or learning an inverse design map is inherently ill-posed, because multiple topologies may have similar effective properties (e.g., the same elastic stiffness can be achieved by different spinodoid designs \cite{KumarEtAl2020_npj}). Conventional ML strategies require training with a distance metric in the output (structural design) space -- defining how ``close'' the structures are to each other.  Yet, the one-to-many nature of property-to-structure maps is prohibitive in defining such a metric and, in turn, precludes the applicability of related ML strategies.

\begin{figure}
\includegraphics[width=\textwidth]{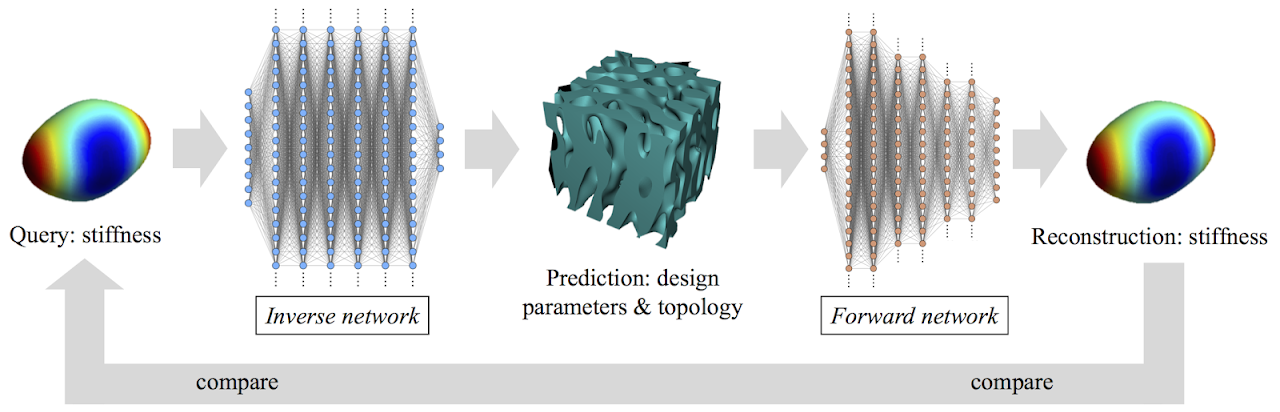}
\caption{Schematic of the inverse design framework \cite{KumarEtAl2020_npj}, which has been applied to the design of spinodoid metamaterials with targeted anisotropic elastic stiffness. The stiffness is visualized via the elasticity surface, wherein each point on the surface denotes Young's modulus in that direction.}
\label{fig:inverse_design}
\end{figure}

To render the inverse design problem well-posed, we introduced an ML approach based on the integration of two neural networks for both the forward and inverse problems \cite{KumarEtAl2020_npj}, which is illustrated in Figure \ref{fig:inverse_design} for the example of spinodoid metamaterials with inverse-designed anisotropic elastic stiffness. The \textit{inverse network} takes the  target stiffness as an input and predicts a design topology. Of course, this predicted topology can be different from the one that was used to generate the training example. Therefore, we do not compare topologies but their effective properties. Since the forward problem is well-posed (each design has unique properties), we use a \textit{forward network} to serve as a surrogate to RVE-based homogenization (pre-trained using simulation data) and reconstructs the stiffness of any predicted design. The inverse network is then trained to minimize the reconstruction error of the target stiffness, and together with the forward network provides a two-way structure-property map.

This on-demand inverse design paradigm is sufficiently general for application to a wide range of (meta-)materials and effective properties. It can, e.g., benefit the design of
patient-specific bone-mimetic scaffolds and implants \cite{KumarEtAl2020_npj} -- bone being highly anisotropic and heterogeneous in its topological and mechanical properties, so that inverse designing bone implants which locally match the native bone properties holds promise for improving the long-term compatibility of bone implants and avoiding atrophy due to property mismatch. 

\section{Conclusions and Outlook}

In this brief contribution, we could only provide a glimpse of the powerful opportunities provided by introducing ML into computational mechanics -- from accelerating multiscale modeling and topology optimization to bypassing constitutive modeling or solving complex systems of PDEs all the way to identifying and inverting structure-property maps. To add a personal perspective, ML in the mechanics community is both underrated and overrated at the same time.  ML-based models are best viewed as powerful and scalable approximators. However, using ML for just about everything is certainly not advised. We are already seeing the limitations of black-box ML methods when applied to physical problems, particularly related to extrapolation, generalization, computational expense, and amount of training data. 
The future lies in the integration of classical physics-based methods with ML methods to address these challenges. By endowing interpretability and insights to data-driven models, we can begin to uncover and understand previously unknown physical phenomena, which have been our pursuits for over a century in mechanics.

\bibliographystyle{spmpsci}
\bibliography{Bib}

\begin{thebibliography}{10}
\providecommand{\url}[1]{{#1}}
\providecommand{\urlprefix}{URL }
\expandafter\ifx\csname urlstyle\endcsname\relax
  \providecommand{\doi}[1]{DOI~\discretionary{}{}{}#1}\else
  \providecommand{\doi}{DOI~\discretionary{}{}{}\begingroup
  \urlstyle{rm}\Url}\fi

\bibitem{BessaEtAl2019}
Bessa, M.A., Glowacki, P., Houlder, M.: Bayesian machine learning in
  metamaterial design: Fragile becomes supercompressible.
\newblock Advanced Materials \textbf{31}(48), 1904845 (2019)

\bibitem{BhattacharyaEtAl2020}
Bhattacharya, K., Hosseini, B., Kovachki, N.B., Stuart, A.: Model reduction and
  neural networks for parametric pdes.
\newblock ArXiv \textbf{abs/2005.03180} (2020)

\bibitem{BruntonEtAl2016}
Brunton, S.L., Proctor, J.L., Kutz, J.N.: Discovering governing equations from
  data by sparse identification of nonlinear dynamical systems.
\newblock Proceedings of the National Academy of Sciences \textbf{113}(15),
  3932--3937 (2016).
\newblock \doi{10.1073/pnas.1517384113}

\bibitem{CapuanoEtAl2019}
Capuano, G., Rimoli, J.J.: Smart finite elements: A novel machine learning
  application.
\newblock Computer Methods in Applied Mechanics and Engineering \textbf{345},
  363 -- 381 (2019).
\newblock \doi{10.1016/j.cma.2018.10.046}

\bibitem{CarraraEtAl2020}
Carrara, P., {De Lorenzis}, L., Stainier, L., Ortiz, M.: Data-driven fracture
  mechanics.
\newblock Computer Methods in Applied Mechanics and Engineering \textbf{372},
  113390 (2020)

\bibitem{CzarneckiEtAl2017}
Czarnecki, W.M., Osindero, S., Jaderberg, M., Swirszcz, G., Pascanu, R.:
  Sobolev training for neural networks.
\newblock CoRR \textbf{abs/1706.04859} (2017)

\bibitem{EggersmannEtAl2019}
Eggersmann, R., Kirchdoerfer, T., Reese, S., Stainier, L., Ortiz, M.:
  Model-free data-driven inelasticity.
\newblock Computer Methods in Applied Mechanics and Engineering \textbf{350},
  81 -- 99 (2019)

\bibitem{FlaschelEtAl2020}
Flaschel, M., Kumar, S., {De Lorenzis}, L.: Unsupervised discovery of
  interpretable hyperelastic constitutive laws.
\newblock Computer Methods in Applied Mechanics and Engineering \textbf{381},
  113852 (2021).
\newblock \doi{10.1016/j.cma.2021.113852}

\bibitem{HuangEtAl2020}
Huang, D., Fuhg, J.N., Wei{\ss}enfels, Wriggers, P.: A machine learning based
  plasticity model using proper orthogonal decomposition.
\newblock Computer Methods in Applied Mechanics and Engineering \textbf{365},
  113008 (2020).
\newblock \doi{10.1016/j.cma.2020.113008}

\bibitem{IbanezEtAl2018}
Iba{\~n}ez, R., Abisset-Chavanne, E., Aguado, J.V., Gonzalez, D., Cueto, E.,
  Chinesta, F.: A manifold learning approach to data-driven computational
  elasticity and inelasticity.
\newblock Archives of Computational Methods in Engineering \textbf{25}(1),
  47--57 (2018)

\bibitem{IyerEtAl2020}
Iyer, A., Zhang, Y., Prasad, A., Gupta, P., Tao, S., Wang, Y., Prabhune, P.,
  Schadler, L.S., Brinson, L.C., Chen, W.: Data centric nanocomposites design
  via mixed-variable bayesian optimization.
\newblock Mol. Syst. Des. Eng. \textbf{5}, 1376--1390 (2020).
\newblock \doi{10.1039/D0ME00079E}

\bibitem{KirchdoerferEtAl2016}
Kirchdoerfer, T., Ortiz, M.: Data-driven computational mechanics.
\newblock Computer Methods in Applied Mechanics and Engineering \textbf{304},
  81 -- 101 (2016).
\newblock \doi{10.1016/j.cma.2016.02.001}

\bibitem{KochmannEtAl2019}
Kochmann, D.M., Hopkins, J.B., Valdevit, L.: Multiscale modeling and
  optimization of the mechanics of hierarchical metamaterials.
\newblock MRS Bulletin \textbf{44}(10), 773–781 (2019).
\newblock \doi{10.1557/mrs.2019.228}

\bibitem{Kramer1991}
Kramer, M.A.: Nonlinear principal component analysis using autoassociative
  neural networks.
\newblock AIChE Journal \textbf{37}(2), 233--243 (1991)

\bibitem{KumarEtAl2020_npj}
Kumar, S., Tan, S., Zheng, L., Kochmann, D.M.: Inverse-designed spinodoid
  metamaterials.
\newblock npj Computational Materials \textbf{6}(1), 73 (2020).
\newblock \doi{10.1038/s41524-020-0341-6}

\bibitem{KumarEtAl2020_cms}
Kumar, S., Tutcuoglu, A.D., Hollenweger, Y., Kochmann, D.: A meshless
  multiscale approach to modeling severe plastic deformation of metals:
  Application to {ECAE} of pure copper.
\newblock Computational Materials Science \textbf{173}, 109329 (2020).
\newblock \doi{10.1016/j.commatsci.2019.109329}

\bibitem{KumarEtAl2020_nme}
Kumar, S., Vidyasagar, A., Kochmann, D.M.: An assessment of numerical
  techniques to find energy-minimizing microstructures associated with
  nonconvex potentials.
\newblock International Journal for Numerical Methods in Engineering
  \textbf{121}(7), 1595--1628 (2020).
\newblock \doi{10.1002/nme.6280}

\bibitem{LiEtAl2020}
Li, Z.Y., Kovachki, N.B., Azizzadenesheli, K., Liu, B., Bhattacharya, K.,
  Stuart, A., Anandkumar, A.: Fourier neural operator for parametric partial
  differential equations.
\newblock ArXiv \textbf{abs/2010.08895} (2020)

\bibitem{MaoEtAl2020}
Mao, Y., He, Q., Zhao, X.: Designing complex architectured materials with
  generative adversarial networks.
\newblock Science Advances \textbf{6}(17), eaaz4169 (2020)

\bibitem{MozaffarEtAl2019}
Mozaffar, M., Bostanabad, R., Chen, W., Ehmann, K., Cao, J., Bessa, M.A.: Deep
  learning predicts path-dependent plasticity.
\newblock Proceedings of the National Academy of Sciences \textbf{116}(52),
  26414--26420 (2019).
\newblock \doi{10.1073/pnas.1911815116}

\bibitem{PandeyEtAl2021}
Pandey, A., Pokharel, R.: Machine learning based surrogate modeling approach
  for mapping crystal deformation in three dimensions.
\newblock Scripta Materialia \textbf{193}, 1--5 (2021)

\bibitem{RaissiEtAl2017_1}
Raissi, M., Perdikaris, P., Karniadakis, G.: Physics informed deep learning
  (part i): Data-driven solutions of nonlinear partial differential equations.
\newblock ArXiv \textbf{abs/1711.10561} (2017)

\bibitem{RaissiEtAl2017_2}
Raissi, M., Perdikaris, P., Karniadakis, G.: Physics informed deep learning
  (part ii): Data-driven discovery of nonlinear partial differential equations.
\newblock ArXiv \textbf{abs/1711.10566} (2017)

\bibitem{RaissiEtAl2019}
Raissi, M., Perdikaris, P., Karniadakis, G.E.: Physics-informed neural
  networks: A deep learning framework for solving forward and inverse problems
  involving nonlinear partial differential equations.
\newblock Journal of Computational Physics \textbf{378}, 686--707 (2019)

\bibitem{ReimannEtAl2019}
Reimann, D., Nidadavolu, K., ul~Hassan, H., Vajragupta, N., Glasmachers, T.,
  Junker, P., Hartmaier, A.: Modeling macroscopic material behavior with
  machine learning algorithms trained by micromechanical simulations.
\newblock Frontiers in Materials \textbf{6}, 181 (2019).
\newblock \doi{10.3389/fmats.2019.00181}

\bibitem{SigmundEtAl2013}
Sigmund, O., Maute, K.: Topology optimization approaches.
\newblock Structural and Multidisciplinary Optimization \textbf{48}(6),
  1031--1055 (2013)

\bibitem{TutcuogluEtAl2020}
Tutcuoglu, A., Hollenweger, Y., Stoy, A., Kochmann, D.: High- vs. low-fidelity
  models for dynamic recrystallization in copper.
\newblock Materialia \textbf{7}, 100411 (2019)

\bibitem{VlassisEtAl2020}
Vlassis, N.N., Ma, R., Sun, W.: Geometric deep learning for computational
  mechanics {Part} {I}: anisotropic hyperelasticity.
\newblock Computer Methods in Applied Mechanics and Engineering \textbf{371},
  113299 (2020).
\newblock \doi{10.1016/j.cma.2020.113299}

\bibitem{WangEtAl2020}
Wang, L., Chan, Y.C., Ahmed, F., Liu, Z., Zhu, P., Chen, W.: Deep generative
  modeling for mechanistic-based learning and design of metamaterial systems.
\newblock Computer Methods in Applied Mechanics and Engineering \textbf{372},
  113377 (2020).
\newblock \doi{10.1016/j.cma.2020.113377}

\bibitem{WhiteEtAl2019}
White, D.A., Arrighi, W.J., Kudo, J., Watts, S.E.: Multiscale topology
  optimization using neural network surrogate models.
\newblock Computer Methods in Applied Mechanics and Engineering \textbf{346},
  1118 -- 1135 (2019).
\newblock \doi{10.1016/j.cma.2018.09.007}

\bibitem{ZhangEtAl2019}
Zhang, Y., Li, H., Xiao, M., Gao, L., Chu, S., Zhang, J.: Concurrent topology
  optimization for cellular structures with nonuniform microstructures based on
  the kriging metamodel.
\newblock Structural and Multidisciplinary Optimization \textbf{59}(4),
  1273--1299 (2019)

\bibitem{ZhengEtAl2020_topopt}
Zheng, L., Kumar, S., Kochmann, D.M.: Data-driven topology optimization of
  spinodoid metamaterials with seamlessly tunable anisotropy.
\newblock Computer Methods in Applied Mechanics and Engineering \textbf{383},
  113894 (2021).
\newblock \doi{10.1016/j.cma.2021.113894}

\bibitem{ZhuEtAl2018}
Zhu, Y., Zabaras, N.: Bayesian deep convolutional encoder-decoder networks for
  surrogate modeling and uncertainty quantification.
\newblock Journal of Computational Physics \textbf{366}, 415 -- 447 (2018).
\newblock \doi{10.1016/j.jcp.2018.04.018}

\end{thebibliography}

\end{document}